\documentclass[useAMS,usegraphicx]{mn2e}

\usepackage[dvips]{color}

\title[Hadronic gamma emission from young radio lobes]
{
Mini radio lobes in AGNs core illumination 
and their hadronic $\gamma$-ray afterlight}
\author[M. Kino, K. Asano]
{Motoki Kino$^{1}$ and
Katsuaki Asano$^{2}$ \\ 
$^{1}$ National Astronomical Observatory of Japan, 181-8588 Mitaka, Japan\\
$^{2}$ Interactive Research Center for Science, Tokyo Institute of Technology,
2-12-1 Ookayama, Tokyo 152-8550, Japan}
\begin{document}

\date{}

\pagerange{\pageref{firstpage}--\pageref{lastpage}} \pubyear{2010}

\maketitle

\label{firstpage}

\begin{abstract}

Recent radio observations reveal the existence of 
mini radio lobes in active galaxies with their scales 
of $\sim 10~{\rm pc}$.
The lobes are expected to be filled with
shock accelerated electrons and protons.
In this work, we examine
the photon spectra from the mini lobes,
properly taking the hadronic processes into account. 
We find that the resultant broadband spectra contain
the two distinct hadronic bumps in $\gamma$-ray bands,
i.e.,  the proton synchrotron bump at $\sim$ MeV and
the synchrotron bump at $\sim$ GeV 
due to the secondary electrons/positrons produced via 
photo-pion cascade. 
Especially when the duration of particle injection is 
shorter than the lobe age,  radio-dark $\gamma$-ray 
lobes are predicted.
The existence of the $\gamma$-ray lobes
could be testable with the future TeV-$\gamma$
telescope {\it CTA}.

\end{abstract}

\section{Introduction}

Thanks to the progress of VLBI
(Very Long Baseline Interferometry) observations, 
compact radio lobes
with a linear size $LS\sim 1~{\rm kpc}$,
defined as a projected length from the core to the lobe,
have been discovered
(e.g., 
   Fanti et al. 1995; 
Readhead et al. 1996;
O'Dea \& Baum 1997).
Further VLBI observations 
recently reveal the existence of very small
radio lobes with $LS\sim 10~{\rm pc}$ 
among two samples of compact radio sources. 
The sample of compact radio 
sources with their spectral peaks higher than 
$\sim5~{\rm GHz}$
is termed as high frequency peakers (HFPs). 
Some of them turn out to be mini lobes 
based on their  morphologies and non-variabilities 
(e.g., Orienti et al. 2007; Orienti \& Dallacasa  2008).
The sample of compact radio sources at 
low-redshift ($z<0.16$) is selected and called as CORALZ. 
The VLBI observations of CORALZ show that some of them 
are also found as mini radio lobes 
(Snellen et al. 2004;  de Vries et al. 2009). 
Apart from the above two cases,
recurrent radio sources are also known to possess mini lobes
inside large lobes (e.g., 3C 84, Walker et al. 2000).
These mini radio lobes are young and their ages are typically
estimated as $t_{\rm age}\sim 10^{2-3}~{\rm yrs}$
(e.g., Fanti 2009, Giroletti \& Polatidis 2009).
High-energy emission of the mini radio lobes have been
theoretically explored by some authors 
(Stawarz et al. 2008; Kino et al. 2007, 2009).
However, previous work has focused on leptonic processes
and little is known about hadronic processes 
in the lobes.

In this Letter,  we indicate that
the hot spots in the mini radio 
lobe can be a plausible site of proton acceleration.
Therefore,  high energy protons are naturally
expected in the mini lobes.
In order to constrain the amount of high energy protons,
next we examine the predicted photon spectra.
Because of their smallness,  observed mini radio lobes 
generally have dense synchrotron photon fields.
Furthermore, the mini lobes are close to the AGN core
and illuminated by the core emission.
Therefore $p\gamma$ interaction is inevitable 
for the mini lobes. 
We calculate the photon spectra from the mini lobes
taking into account the hadronic processes and
show that the spectra are useful for 
constraining the amount of protons 
via $\gamma$-ray emission features.
The predicted spectra will be useful for testing 
whether cosmic ray acceleration indeed takes place in mini lobes.


\section{Hadronic emission}

When a jet interacts with the surrounding medium,
most of its kinetic energy is dissipated via shocks.
The termination point of the jet 
is known as a hot spot and it is identified as the reverse
shocked region. The shocked jet plasma 
escapes sideways from the spots 
forming a pair of radio lobes.
Therefore, radio lobes are 
remnants of the decelerated jets 
which contain relativistic particles 
(e.g., Begelman et al. 1984).
As for a ratio of electrons to protons in AGN jets, 
it is still under debate
(e.g.,  Kino and Takahara (2004), Birzan et al. (2008)).
There are case studies indicating that protons 
are dynamically dominant at hot spots of radio galaxies 
(e.g., Stawarz et al. 2007).
Here we examine what happens 
when assuming the lobe contains relativistic protons. 
The notation $Q=Q_{x}\times 10^{x}$ is used 
unless otherwise noted.

\subsection{Proton acceleration at the hot spots}

Hot spots in powerful radio galaxies are one of 
the most promising sites for proton acceleration 
(e.g., Rachen and Biermann 1993). 
Here we show that hot spots in mini lobes are also
possible sights for proton acceleration.
The energy of shock accelerated protons is given by
$\varepsilon_{p}=\gamma_{p}m_{p}c^{2}$
where $\gamma_{p}$ is the proton Lorentz factor measured in the 
shock frame.
Although a Bohm-type diffusion timescale is a
rough approximation, we adopt it for simplicity.
Then the acceleration timescale 
at the  hot spot 
$t_{p,\rm acc,hs}=\xi_{p} \varepsilon_{p}/(eB_{\rm  hs}c)$ 
in the case of relativistic shocks can be estimated as
\begin{eqnarray}
t_{p,\rm acc,hs}
\approx 3.5~\varepsilon_{p,18}B_{\rm hs, -1}^{-1}\xi_{p,2}~{\rm yr} ,
\end{eqnarray}
where  
$B_{\rm hs}$, and
$\xi_{p}$ are the magnetic field strength, and
the Gyro-factor for proton accelerations 
at the hot spot, respectively.
Although there is no direct estimate of $B_{\rm hs}$,
magnetic field strength in  mini lobes ($B_{\rm lobe}$) 
has recently been estimated as
$B_{\rm lobe} \sim 10-100 ~{\rm mG}$ 
by VLBI observations (Orienti \& Dallacasa 2008). 
Since the magnetic field strength 
at the hot spot $B_{\rm hs}$ should be comparable to
or larger than  
$B_{\rm lobe}$,
here we set $B_{\rm hs}\sim 100~{\rm mG}$.
The value of $\xi_{p}$ is a free parameter.
In the case of blazars, 
$\xi_{p}$ is not expected to be smaller than $10$
(Inoue and Takahara 1996).
As for mini lobes, there is little constraint on  $\xi_{p}$. 
As a first step,
we only treat the case of $\xi_{p}= 10^{2}$ in this Letter.
The case of  $\xi_{p}<10^{2}$ is not observationally excluded
and it is worth to examine it in the context of highest energy 
cosmic ray sources.
When $\xi_{p}\gg 10^{2}$, photo-pion cascade does
not occur. So, we do not treat it.
The timescale of proton synchrotron cooling is
\begin{eqnarray}\label{eq:t_syn}
t_{p,\rm syn,hs}=
\frac{6\pi m_{p}^{4}c^{3}}{\sigma_{T}m_{e}^{2}\varepsilon_{p}B_{\rm hs}^{2}}
\approx 1.4 \times 10^{4}~
\varepsilon_{p,18}^{-1}B_{\rm hs,-1}^{-2}~{\rm yr} ,
\end{eqnarray}
where $\sigma_{\rm T}$ is the Thomson cross section.
The escape velocity of shocked matter from the hot spot
via sideways expansions is typically
$v_{\rm esc, hs}\approx 0.3~c$ (e.g., Kino and Takahara 2004).
Then the escape timescale is 
$t_{\rm esc,hs}\approx 3~R_{\rm hs,18}~{\rm years}$
where $R_{\rm hs}$ is the hot spot radius. 
Since $t_{\rm esc,hs} \ll t_{\rm syn,hs}$
is satisfied, the maximum energy of 
protons $\varepsilon_{p,\rm max}$  is obtained by
the relation  
$t_{p,\rm acc,hs}= t_{\rm esc,hs}$ and it
is 
\begin{eqnarray}
\varepsilon_{p,\rm max}\approx 
0.9 \times 10^{18} B_{\rm hs,-1}
R_{\rm hs, 18}\xi_{p,2}^{-1}
~{\rm eV} .
\end{eqnarray}
%
The Larmor radius of protons in the spot ($r_{\rm L,hs}$) 
satisfies  $R_{\rm hs}>r_{\rm L,hs}
\approx 1\times 10^{-2}~\varepsilon_{p,18}
B_{\rm hs,-1}^{-1}~{\rm pc}$.
Thus we find that 
the hot spots in mini lobes are feasible  sites
for proton acceleration.
%
%
The escaped protons from the spots 
are then injected into the mini lobes and
subsequently undergo cooling there.

\subsection{Proton cooling in the mini radio lobes}

The relativistic protons injected in  mini lobes
undergo various coolings. 
Before we describe the results of the Monte Carlo
simulation in detail,
it is worth estimating the relevant cooling timescales.
First, the adiabatic loss timescale at the mini lobes is given by
$t_{\rm ad}
\approx 65~(R_{\rm lobe}/2~{\rm pc})
(v_{\rm exp,lobe}/0.1c)^{-1}~{\rm yrs}$ 
where 
$R_{\rm lobe}$ and $v_{\rm exp,lobe}$
are the radius and expansion velocity of the lobe,  respectively.
As  shown above, the Larmor radius of protons of $10^{18}~{\rm eV}$
is sufficiently smaller than $R_{\rm lobe}$. 
Therefore, they are confined in the lobes
and  suffer adiabatic expansion loss.
Next,
we estimate the timescale of 
the photo-meson loss process of high energy protons.
As shown in the introduction, 
$p\gamma$ interactions are expected in the lobe
where target photons are UV photons from the accretion disk
and synchrotron photons emitted by the lobe itself.
%
%
The condition to create pions is 
$\varepsilon_{p}\varepsilon_{\gamma,i} \ge 0.2~{\rm GeV}^{2}$
where $\varepsilon_{\gamma,i}$ is the target photon energy 
($i=\rm disk, syn$).
Then, the typical energy of protons 
interacting with the accretion disk photons is given by
$\varepsilon_{p} \approx 
2\times 10^{16}(\varepsilon_{\gamma,{\rm disk}}/10~{\rm eV})^{-1}~{\rm eV}$.
The timescale of the $p\gamma$ interaction is
$t_{p\gamma}\approx
(n_{\gamma,i}\sigma_{p\gamma} c)^{-1}
\approx 1.2 \times 10^{4}~
(LS/10~{\rm pc})^{-2}
L_{\rm disk,45}
(\varepsilon_{\gamma,{\rm disk}}/10~{\rm eV})^{-1}~{\rm yr}$
where 
$\sigma_{p\gamma}\approx 5\times 10^{-28}~{\rm cm^{-2}}$ 
at the resonance peak (Waxman and Bahcall 1997)
and 
$L_{\rm disk}$ and
$n_{\gamma,{\rm disk}}$ are
the disk luminosity and
the photon number density  
$n_{\gamma,{\rm disk}}
\approx
L_{\rm disk}/(4\pi LS^{2}c \varepsilon_{\gamma,{\rm disk}})$, respectively.
For $L_{\rm disk}$, we
assume a value of observed big blue bumps 
in Seyfert 2 galaxies (e.g., Koratkar and Blaes 1999).
In this case, we have
$t_{\rm ad}/t_{p \gamma}\approx 5.3\times 10^{-3}
R_{\rm lobe}LS^{2}L_{\rm disk}^{-1}
\propto R_{\rm lobe}^{3}$ 
when we assume $LS\propto R_{\rm lobe}$.
For protons with $\varepsilon_{p}>2\times 10^{16}~{\rm eV}$,
the target photons are  the electron synchrotron ones.
In this range,
$n_{\gamma,{\rm syn}}\approx
L_{\rm syn}/(\pi R_{\rm lobe}^{2}c \varepsilon_{\gamma,{\rm syn}})
\approx 1.7 \times 10^{4}
~L_{\rm syn, 44} (R_{\rm lobe}/2~{\rm pc})^{-2}
(\varepsilon_{\gamma,{\rm syn}}/10^{-6}~{\rm eV})^{-1}
~{\rm cm^{-3}}$
where $L_{\rm syn}$ is the synchrotron luminosity emitted by
primary electrons.
Then we have
$t_{\rm ad}/t_{p \gamma}
\approx 5.3\times 10^{-4}~R_{\rm lobe}^{3}L_{\rm syn,44}^{-1}
(\varepsilon_{\gamma, {\rm syn}}/10^{-6}~{\rm eV})^{-1}
\propto R_{\rm lobe}^{3}$.
So, $t_{p \gamma}$ become shorter than $t_{\rm ad}$
when $R_{\rm lobe}$ is sufficiently small, although such a case 
is beyond the scope of this work.
The timescale of proton synchrotron loss at the lobe
has been already given by Eq.~(\ref{eq:t_syn})
since we assume $B_{\rm lobe}\approx B_{\rm hs}$
for simplicity.

The timescale of energy loss by proton-proton collisions 
is given by  
$t_{pp}\sim 5\times 10^{7}(n_{\rm ext}/1~{\rm cm^{-3}})^{-1}~{\rm yr}$
(Sikora et al. 1987)
where $n_{\rm ext}$ is the number density of the ambient matter.
Here we focus on the case of $t_{pp}\gg t_{\rm age}\sim 10^{2-3}~{\rm yr}$.
However, there is a possibility that mini lobes expand 
in rather dense environments involving clouds of gas emitting 
narrow emission lines with $n_{\rm ext}\gg 1~{\rm cm^{-3}}$,
and that $t_{pp}\sim t_{\rm age}$ holds. 
We will examine it in our future work.


\section{Resultant spectra}

To calculate broadband photon spectra, 
the Monte Carlo simulation has been performed 
in this work. The details of numerical code have been
described in Asano et al. (2007, 2008, 2009) and references therein.
Here we briefly review  the code.
We calculate  steady state spectra of 
protons ($i=p$) and electrons ($i=e$)
by solving the following kinetic equations
\begin{eqnarray}\label{kin-eq}
q_{i}(\varepsilon_{i}) = -\frac{\partial}{\partial \varepsilon_{i}}
\left[n_{i}(\varepsilon_{i}){\dot \varepsilon_{i}}\right]+ 
\frac{n_{i}(\varepsilon_{i})}{t_{\rm ad}},
\end{eqnarray}
where 
$n_{i}(\varepsilon_{i})$, 
$\dot{\varepsilon}_{i}$, 
and
$q_{i}(\varepsilon_{i}) = K_{i} \varepsilon_{i}^{-s_{i}} 
\exp\left(\frac{\varepsilon_{i}}{\varepsilon_{i,\rm max}}\right)$
for
$\varepsilon_{i,{\rm min}}\le \varepsilon_{i}$ 
are
the number density,
the cooling rate, and
the injection rate of relativistic particles, 
respectively. 
Distributions of particles and photons are assumed to be isotropic.
We assume that the common Fermi acceleration process works 
both for electrons and protons.
In this case,
the response to magnetic turbulences is same
for relativistic particles with a same energy.
Therefore, 
it is conservative 
to set  $s_{e}=s_{p}$ and  $\xi_{e}=\xi_{p}$.
As explained in 2.1,
$\varepsilon_{p,\rm max}$ are uniquely
determined for given hot spot parameters.
The maximum energy of electrons 
can be obtained as
$\varepsilon_{e,\rm max}
\sim 10~B_{\rm hs, -1}^{-1/2}
\xi_{e,2}^{-1/2}~{\rm TeV}$
from the relation $t_{e,\rm syn,hs}=t_{e,\rm acc, hs}$ where
$t_{e,\rm syn,hs}$ and
$t_{e,\rm acc,hs}=(\xi_{e}\varepsilon_{e})/eB_{\rm hs}c$ are the
synchrotron cooling and acceleration 
timescales for electrons, respectively.
The kinetic equation of photons 
with the energy $\varepsilon_{\gamma}$ is given by
\begin{eqnarray}\label{photon-kin-eq}
\frac{\dot{N}_{\gamma}(\varepsilon_{\gamma})}{\pi R_{\rm lobe}^{2}c}=
n_{\gamma}(\varepsilon_{\gamma}),
\end{eqnarray}
where $\dot{N}_{\gamma}(\varepsilon_{\gamma})$ 
is  the total production rate of each population of photons and
$n_{\gamma}(\varepsilon_{\gamma})$ is the corresponding
photon number density.
As for $\dot{N}_{\gamma}$ and 
${\dot \gamma_{p}}$,
we include the following physical processes:
(1) photo-pion production from protons and neutrons 
($p + \gamma \rightarrow p/n + \pi^{0}/\pi^{+}$),
(2)
the pions decay ($\pi^{0} \rightarrow 2 \gamma$) and 
($\pi^{+} \rightarrow \mu^{+} + \nu_{\mu}
\rightarrow e^{+} + \nu_{e}+ \nu_{\mu}+ \bar{\nu}_{\mu}$), 
(3)
photon-photon pair production 
($\gamma + \gamma \rightarrow e^{+} + e^{-}$),
(4) 
Bethe-Heitler pair production 
($p + \gamma \rightarrow p + e^{+} + e^{-}$),
(5)
synchrotron and inverse Compton processes 
from electrons/positrons,
protons, pions, muons with Klein-Nishina cross section, and
(6)
synchrotron self-absorption for electrons/positrons.

The model parameter values are summarized here.
For the hot spot, we set
$R_{\rm hs}=0.3~{\rm pc}$,
$B_{\rm hs}=0.1~{\rm G}$, and
$v_{\rm esc,hs}=c/3$. 
These hot spot quantities
uniquely determine $\varepsilon_{i,{\rm max}}$.
We adopt
$\xi_{p}=\xi_{e}=1\times 10^2$,
$s_{p}=s_{e}=2$, and
$\varepsilon_{p,{\rm min}}/m_{p}c^{2}
=\varepsilon_{e,{\rm min}}/m_{e}c^{2}=10$ 
for particle injections. 
The mini lobes parameters are
$R_{\rm lobe}=2~{\rm pc}$,
$B_{\rm lobe}=0.1~{\rm G}$, 
$v_{\rm exp,lobe}=0.1~c$, and
$LS=10~{\rm pc}$.
The accretion disk quantities are assumed as
$L_{\rm disk}=3 \times 10^{45}~{\rm erg~s^{-1}}$
with the average temperature $10~{\rm eV}$.
Because of the obscuration by surrounding dusty torus
$L_{\rm disk}$ in mini radio lobes is not directly constrained.
Following the work of Ostorero et al. (2010),
here we assume $L_{\rm disk}$ as the same one 
in luminous blazars.
The injection power of protons $L_{p}$ should be
smaller than the total kinetic powers of powerful
jets $\sim 10^{47-48}~{\rm erg~s^{-1}}$ 
(e.g., Ito et al. 2008, Ghisellini et al. 2009). Here
$ L_{p}=5\times 10^{46}~{\rm erg~s^{-1}}$ is assumed.
Various injection powers of electrons $L_{e}$
will be examined below. 


\subsection{Leptonic-hadronic model}

Fig. 1 displays the resultant photon spectrum for
$L_{e}=1\times 10^{42}~{\rm erg~s^{-1}}$ (the thick solid line).
There are several distinct features in the spectrum.
\begin{enumerate}
\item
The bump at $\sim~{\rm sub~PeV}$ energies
is composed of $\pi^{0}$-decay photons.
The $\gamma\gamma$ absorption opacity
against the target photons with the energy
$\varepsilon_{\gamma,{\rm tgt}}$ 
at the threshold 
$\varepsilon_{\gamma} 
\varepsilon_{\gamma,{\rm tgt}} \sim (2m_{e}c^{2})^{2}$
is written as 
$\tau_{\gamma \gamma}(\varepsilon_{\gamma}) 
=(3/8)\sigma_{\rm T} n_{\gamma,{\rm tgt}}R_{\rm lobe}$ 
(e.g., Razzaque et al. 2004) and it is estimated as
\begin{eqnarray} \label{eq:tau_gg}
\tau_{\gamma \gamma}(\varepsilon_{\gamma}=10^{14}~{\rm eV}) 
&\approx& 
0.8~ 
\frac{L_{\gamma,{\rm tgt}}}{3\times 10^{40}~{\rm erg~s^{-1}}}
\frac{R_{\rm lobe}}{2~{\rm pc}} \nonumber \\
&&\times
\left(\frac{\varepsilon_{\gamma,{\rm tgt}}} 
{10^{-2}~{\rm eV}}\right)^{-1}
\end{eqnarray}
where 
the number density of target photons is
$n_{\gamma, {\rm tgt}} =
\dot{N}_{\gamma,{\rm tgt}}/(\pi R_{\rm lobe}^{2}c)=
L_{\gamma,{\rm tgt}}/(\pi  \varepsilon_{\gamma,{\rm tgt}}R_{\rm lobe}^{2}c)$.
Hence,  the $\pi^{0}$-decay photons with
$\varepsilon_{\gamma}\sim 10^{14}~{\rm eV}$ 
can partially escape from the lobe.
The escaped $100~{\rm TeV}$ photons, however, 
would be absorbed  
by the radio background photons during their propagations
since their mean free path is shorter than $10~{\rm Mpc}$
(e.g., Coppi and Aharonian 1997).
%

\item
The secondary electrons/positrons produce the 
synchrotron bump at  $\sim $ GeV.
They are  produced via
$\mu$- and  $\pi$-decays.
The $e^{\pm}$ pairs created by 
the $\gamma\gamma$ absorption of 
the $\pi^{0}\rightarrow 2\gamma$ photons
have an energy  
of $1/2 \times 1/2=1/4$ that of the parent $\pi^{0}$.
In the $\pi^{+}$ decay mode, 
the comparable fraction of the energy is converted from
the parent $\pi^{+}$ to the positron.
Then, the synchrotron emission from
secondary electrons/positrons with $\sim 10^{15}~{\rm eV}$
has its peak at  $\sim $ GeV.

\item
The proton synchrotron bump (p-SYN)
appears  at MeV energy band
with the peak at
$\nu_{p,\rm syn}\approx
(\varepsilon_{p}^{2}eB_{\rm lobe}/2\pi m_{p}^{3}c^{5}) \approx
0.7~\varepsilon_{p,18}^{2}B_{{\rm lobe},-1}~{\rm MeV}$. 
%
Below $\sim {\rm keV}$, 
the synchrotron emission from the primary electrons
overwhelms the proton synchrotron emission.
Protons also transfer energy to photons by 
inverse Compton scattering
and the photon spectrum is shown in 
the thin line (p-IC). 
However, they do not escape from the lobe because of the
$\gamma\gamma$ absorption.
\item
The break of synchrotron spectrum at $\sim {\rm GHz}$ 
is well known  synchrotron self absorption (SSA)
observed by VLBI observations (e.g., Snellen et al. 2000).
The SSA turnover frequency 
for $s_{e}=2$ is given by 
$\nu_{\rm ssa}
\sim
0.39~B_{{\rm lobe},-1}^{1/7}
\left(
\frac{L_{\rm ssa}}{10^{41}~{\rm erg~s^{-1}}}\right)^{2/7}
\left(
\frac{R_{\rm lobe}}{2~{\rm pc}}\right)^{-4/7} 
~{\rm GHz}$ 
(Kellermann and Pauliny-Toth 1981)
where
$L_{\rm syn}>L_{\rm ssa}\equiv \nu_{\rm ssa}L_{\nu_{\rm ssa}}$.
The electrons with $\gamma_{e}\sim 100$ emitting 
synchrotron at $\sim 1~{\rm GHz}$ will cool down,
since the synchrotron cooling timescale satisfies
$t_{e,\rm syn}
\sim 25~B_{{\rm lobe},-1}(\gamma_{e}/10^{2})^{2}{\rm yr}
<t_{\rm ad}$.
Therefore the cooling 
break frequency obtained 
from $t_{e,\rm syn}=t_{\rm ad}$ is
comparable to or below  
$\nu_{\rm ssa}$. 
This implies that $L_{e}\approx L_{\rm syn}$.

\item
For comparison,
the case of no-proton injection 
(i.e., $L_{p}=0$)
is shown in the thin solid line. 
By comparing the spectrum to the 
one with proton injection, 
we can clearly recognize the
contribution of the hadronic emission 
in the $\gamma$-ray energy domain.

Fig. 2 shows the photon spectra for
$ L_{e}=1\times 10^{45}~{\rm erg~s^{-1}}$, 
$ 1\times 10^{44}~{\rm erg~s^{-1}}$, 
$ 1\times 10^{43}~{\rm erg~s^{-1}}$, 
$ 1\times 10^{42}~{\rm erg~s^{-1}}$, and 
$ 1\times 10^{41}~{\rm erg~s^{-1}}$.
The $L_{p}/L_{e}$ ratio in AGN jets is not well known.
The examined $L_{p}/L_{e}$ ratio here
partly exceeds the ratio of $\sim 10^{2}$ measured around the Earth.
However, when a jet with $ L_{p}/L_{e}\sim 10^{2}$ gradually 
decreases its power keeping the $ L_{p}/L_{e}$ ratio,
the resultant spectrum may be similar to 
the one for larger $ L_{p}/L_{e}$
because of fast cooling of electrons.
The spectra with and without proton injection
are displayed with thick and thin solid lines, respectively.
For $L_{e}=1\times 10^{45}~{\rm erg~s^{-1}}$,
the leptonic emissions 
overwhelm hadronic ones at all energy domains.
As $L_{e}$ 
decreases to $L_{e}\le 1\times 10^{44}~{\rm erg~s^{-1}}$,
thick and thin lines become separable in the $\gamma$-ray domain
because the synchrotron emission from the secondary electrons/positrons 
overwhelms the inverse Compton component from primary accelerated electrons.
In MeV range,
the proton synchrotron bump appears for smaller $L_{e}$. 
Typical $L_{\rm syn}$ depends on the source population.
CORALZs typically have 
$L_{\rm ssa}\sim 10^{40-42}~{\rm erg~s^{-1}}$
while HFPs 
have $L_{\rm ssa}\sim 10^{44-46}~{\rm erg~s^{-1}}$.
Observed turnover frequencies of mini lobes
shows $\nu_{\rm ssa} \sim 0.1-10~{\rm GHz}$.
The maximum $\nu_{\rm ssa}$ in Fig. 2 is
smaller than the ones for HFPs, mainly because 
the angular size of the lobe in our model is larger
by a factor of a few than the one for HFPs.
It does not affect the main results of this work.
The synchrotron self Compton component appeared
in $\gamma$-ray bands is less dominant than the synchrotron one 
since $U_{\rm syn} \le B_{\rm lobe}^{2}/8\pi\approx 
4\times 10^{-4}B_{{\rm lobe},-1}^{2}~{\rm erg~cm^{-3}}$ holds
for $L_{\rm syn} < 10^{45}~{\rm erg~s^{-1}}$
where  $U_{\rm syn}$ is the energy density of synchrotron photons.

When $L_{e}<10^{44}~{\rm erg~s^{-1}}$,
the TeV photons in the lobe become 
transparent against $\gamma\gamma$
absorption because of
$\tau_{\gamma \gamma}
(\varepsilon_{\gamma}=10^{12}~{\rm eV})\sim 
0.8(L_{\gamma,{\rm tgt}}/3\times 10^{42}~{\rm erg~s^{-1}})
(\varepsilon_{\gamma,{\rm tgt}}/1~{\rm eV})$.
Then the TeV photons begin to escape from the lobe.

%

\end{enumerate}


\subsection{Pure hadronic model}

In Fig. 3, we show the spectrum with
the same parameters as Fig. 1 
but for pure proton injection (i.e., $L_{e}=0$)
with the proton injection duration 
$t_{\rm inj}=90~{\rm yrs}$.
In order to realize $L_{e}=0$ approximation,
$t_{\rm inj}<t_{\rm age}$ and 
$t_{e,\rm syn}<t_{\rm age}$ should be satisfied
where $t_{e,\rm syn}$ is the 
synchrotron cooling timescale for electrons. 
In this case, the injection has been already stopped 
and primary electrons have been already cooled down.
Since the source age in this model can be estimated as
$t_{\rm age}\sim LS/v_{\rm exp}
\sim 3\times 10^{2}~{\rm yr}$
and it agrees with the observational estimates of
$t_{\rm age}\sim 10^{2-3}~{\rm yr}$ (e.g., Giroletti 2009),
the conditions of 
$t_{\rm inj}<t_{\rm age}$ and 
$t_{\rm age}>t_{e,\rm syn}
\sim 25~B_{{\rm lobe},-1}(\gamma_{e}/10^{2})^{2}~{\rm yr}$ 
are indeed justified. 
%
The model considered here is not applicable when $B$ satisfy
$t_{\rm age}<t_{e,\rm syn}$. 
Short $t_{\rm inj}$ would be naturally
realized for jets with intermittent activities.
For example, the recurrent mini lobe 3C 84 actually show 
$t_{\rm inj}\sim 50~{\rm yr}$ (Asada et al. 2006).
The low power mini lobes  
which seem to be dying ones (Giroletti et al. 2005)
could also be candidates.
In this model,
the deposited energy of protons is given by $L_{p}t_{\rm inj}$.

The prominent three bumps of hadronic emissions, i.e.,  
the $\pi^{0}$-decay photon bump at $\sim$ PeV,
the synchrotron one from secondary electrons/positrons 
at $\sim$ GeV, and 
the proton synchrotron one at $\sim$ MeV, 
which have also been predicted in Fig. 1
emerge more clearly.
The escaped PeV photons from the lobe are 
absorbed by the radio background photons 
and they may not reach  the Earth.

Below the optical energy band, 
the synchrotron emission from secondary 
$e^{\pm}$ pairs via Bethe-Heitler process
($p\gamma\rightarrow pe^{-}e^{+}$)
is responsible for the emission.
The protons with
$\varepsilon_{p}\sim 10^{14}~{\rm eV}$
interact with the disk photons and convert 
a fraction of $\sim {m_{e}/m_{p}}$ of their energy
to create $e^{\pm}$ pairs.
Then the pairs with $\sim 10^{11}~{\rm eV}$ 
radiate synchrotron emission peaked  at $\sim$ optical band.
Since the lobes are dim in radio band, we may  
call them radio-dark $\gamma$-ray lobes.

\section{Summary and Discussion}

In this work, we point out that
the hot spots in mini lobes are feasible sites
of proton acceleration. Next,   
the expected photon spectra of the mini lobes 
including the hadronic processes are explored. 
Summary and discussions are as follows.

\begin{enumerate}

\item

For bright lobes with 
$L_{e}\sim L_{\rm syn}\sim 10^{45}~{\rm erg~s^{-1}}$, 
the predicted high energy emission is detectable with the current
$\gamma$-ray telescopes.
However, it is overwhelmed by the leptonic inverse Compton component,
and it seems hard to test whether the emission is of
hadronic- or leptonic-origin. 
For $L_{e}\sim L_{\rm syn}\le 10^{43}~{\rm erg~s^{-1}}$, 
proton synchrotron bump appears  at $\sim$  MeV and 
the synchrotron emission from the secondary  electrons/positrons
generated by the $\mu$- and $\pi$- decays emerges at $\sim$ GeV.
The two distinctive bumps in $\gamma$-ray domain 
are of hadronic origin.

\item

The case of the short term particle injection is examined.
It may be realized when jets have intermittent activities.
Typically, primary accelerated electrons 
have been cooled down but protons have not.
Then, the predicted emission is purely hadronic
and the hadronic bumps are clearly seen.
Importantly,
the high energy tail of the GeV bump is detectable by 
{\it The Cherenkov Telescope Array (CTA)}
(http://www.cta-observatory.org/).
These sources
may be identified as radio dark $\gamma$-ray lobes.


\item

The predicted X-ray flux is well above the detection limit of 
$XMM$. The observations actually show bright $X$-ray emission.
This is traditionally interpreted as thermal radiation from
the accretion disk and the possibility of lobe emission 
has been alternatively
indicated (Ostorero et al. 2010 and reference therein).
In any case, X-ray emission is likely composed of 
various different components.
Therefore, it seems difficult to extract the  
hadronic component from the X-ray band.

\item

We comment on the importance of larger $n_{\rm ext}$
it may lead to frre-free absorption (FFA) in radio band
(e.g., Begelman 1999; Bicknell 2003; Stawarz et al. 2008).
Actually, the low frequency turnover in the radio spectra of some 
sources are indeed reproduced by FFA and not by SSA
(e.g., OQ 208, Kameno et al. 2000; 0108+388, Marr et al. 2001).
It is clear that dense emvironments lead to
an effective proton-proton collision. 
We will examine it in the future.

\item

We add a  comment on the 
recent VLBI observation of mini lobe 3C 84. 
It shows the outburst around 2005 and 
a new component 
smaller than $1~{\rm pc}$ emerges (Nagai et al. 2010). 
Fermi/LAT also detect 
GeV $\gamma$-ray emission from it (Abdo et al. 2009).
A future collaboration with Space VLBI project VSOP-2 
with high angular resolution 
(http://www.vsop.isas.jaxa.jp/vsop2/) and {\it CTA} would provide us 
valuable constraints on the hadronic model.
Theoretically, we plan to conduct studies with smaller $R_{\rm lobe}$ 
in our future work.

\section*{Acknowledgments}

We thank the referee for comments to improve this paper.
We are indebted to 
H. Takami, H. Nagai, M. Orienti and N. Kawakatu
for useful comments and discussions.

\end{enumerate}


\begin{figure}
\includegraphics[width=8cm
]{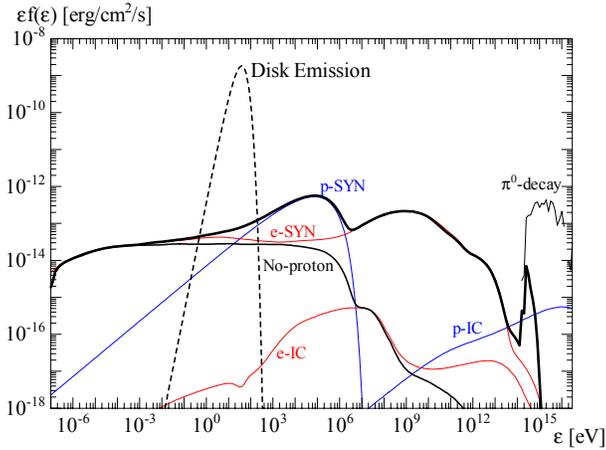}
\caption
{
Predicted photon spectrum
from the mini radio lobe for
$L_{e}=1\times 10^{42}~{\rm erg~s^{-1}}$,
$L_{p}=5\times 10^{46}~{\rm erg~s^{-1}}$ and
$L_{\rm disk}=3\times 10^{45}~{\rm erg~s^{-1}}$.
The thick solid line represents the spectrum
including the effect of internal $\gamma\gamma$ absorption. 
Thin blue lines show the proton emissions, 
while thin red lines represent electron emissions
before including the  $\gamma\gamma$ absorption.
The luminosity is converted to the flux
with the distance $100~{\rm Mpc}$.}
\label{fig:}
\end{figure}
\begin{figure}
\includegraphics[width=8cm
]{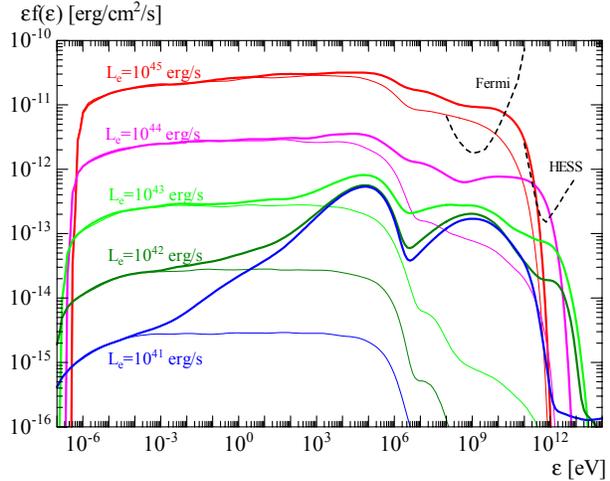}
\caption
{
Same as Fig. 1 but for various 
$L_{\rm e}$
from
$L_{\rm e}=1\times 10^{45}~{\rm erg~s^{-1}}$
to
$1\times 10^{41}~{\rm erg~s^{-1}}$.
The sensitivities of {\it Fermi}/LAT, and HESS are,
respectively, the ones for 1 year, and 50 h integration time.}
\label{fig:}
\end{figure}
\begin{figure}
\includegraphics[width=8cm
]{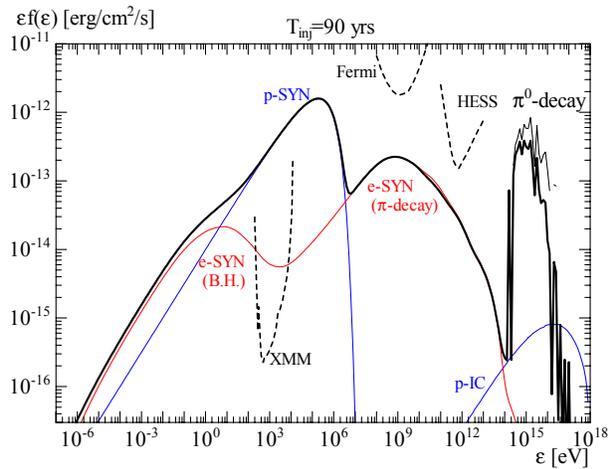}
\caption
{
Same as Fig. 1 but for the case pure proton injection
$t_{\rm inj}=90~{\rm years}$. 
The synchrotron emission 
from electrons/protons produced via Bethe-Heitler process
and pion- and muon-decay are predicted. 
The Proton synchrotron peaks at $\sim$  MeV. 
PeV photons via pion-decay
are absorbed during their propagations.
The sensitivity for 
{\it XMM-Newton} is normalized by  100~ks integration time.}
\label{fig:}
\end{figure}

\end{document}